\documentclass[aps,superscriptaddress,reprint]{revtex4-1}
\usepackage{graphicx}
\usepackage{amsmath}

\begin{document}

\title{Nonlinear dynamics of ionization stabilization of atoms in intense laser fields} 

\author{M.J. Norman}
\affiliation{School of Physics, Georgia Institute of Technology, Atlanta, Georgia 30332-0430, USA}

\author{C. Chandre}
\affiliation{Centre de Physique Th\'eorique, CNRS--Aix-Marseille Universit\'e, Campus de Luminy, 13009 Marseille, France}

\author{T. Uzer}
\affiliation{School of Physics, Georgia Institute of Technology, Atlanta, Georgia 30332-0430, USA}

\author{P. Wang}
\affiliation{The Beijing Key Laboratory for Nano-Photonics and Nano-Structure, Department of Physics, Capital Normal University, Beijing 100048, China}

\date{\today}

\begin{abstract}
We revisit the stabilization of ionization of atoms subjected to a superintense laser pulse using nonlinear dynamics. We provide an explanation for the lack of complete ionization at high intensity and for the decrease of the ionization probability as intensity is increased. We investigate the role of each part of the laser pulse (ramp-up, plateau, ramp-down) in this process. We emphasize the role of the choice for the ionization criterion, energy versus distance criterion. 
\end{abstract}

\pacs{32.80.Rm, 05.45.Ac, 32.80.Fb}

\maketitle 

\section{Introduction}
``Stabilization'' is the term used for the counterintuitive behavior of atoms in which increasing the laser intensity does not lead to increased ionization -- on the contrary, it may lead to decreased ionization in some regimes. Stabilization has been the subject of so many publications that even a cursory review is beyond the scope of our manuscript (for some early references, see Refs.~\cite{gelt74,gers76,gavr84,su90,pont90,grob91,kula91,eber93}, or Ref.~\cite{gavr02} for a review). 

The purpose of our manuscript is to provide new insights into the stabilization phenomenon, gained by viewing it through the unconventional point of view of nonlinear dynamics. This is not to say that the use of classical mechanics to examine stabilization is a novelty: A number of stimulating studies were performed during the time frame mentioned above (e.g., Ref.~\cite{grob91}). However, the intervening two decades have seen the realization of lasers with intensities far beyond what was expected at the time, and quite a few phenomena such as recollisions~\cite{kuch87,cork93,scha93} have been discovered, requiring the inception of various theoretical models and techniques for understanding ionization (or the lack of it) in intense laser fields. 

In the intervening decades laser pulses have become ultrashort (down to about 100 attoseconds). The general idea is to go down to the timescale of the electron to capture its dynamics. Such ultrashort timescales were not part of stabilization research at the time, which typically considered fairly long pulses of several tens of laser cycles. Can any insights about such a long-pulse phenomenon be relevant for understanding the motion of the electron on its timescale? As we will show below, no matter how long the pulse is, the fate of the electron, and hence stabilization, is sealed quite early in the pulse, and specifically during a few first few  laser cycles. In that sense, stabilization turns out to be a short-timescale phenomenon. 

In what follows we will be answering two questions: First, what accounts for the lack of ionization in ultra-intense laser fields, and secondly, what causes ionization to decrease with increasing intensity? It turns out that the answers to these questions, and the hidden short-time nature of the phenomenon, are governed by the same underlying mechanism, namely the periodic orbits of the atomic electron in the laser field and the phase-space structures associated with these periodic orbits. We will show that electron trajectories which can reach the vicinity of the periodic orbit behave completely differently from those trajectories that fail to do so, thereby accounting for the observations connected with stabilization. 

In the process of answering these questions we will also address some highly practical aspects of stabilization, namely~:

1)	How does one decide whether an atomic electron has ionized in an intense laser field? More specifically, are the so-called distance and energy criteria for ionization interchangeable?

2)	What role do the ramp-up, plateau and ramp-down of the laser field play in stabilization?

In Sec.~\ref{sec:IP} we start by performing the numerical experiment of Ref.~\cite{grob91} and assess the importance of the choice of the ionization criterion. In Sec.~\ref{sec:NLanalysis} we provide an explanation for the lack of complete ionization at high intensity and for the decrease of the ionization probability as the laser intensity is increased. This explanation is based on a periodic orbit analysis of the electronic dynamics. We investigate the role of each part of the laser pulse, namely the ramp-up, the plateau and the ramp-down, in the stabilization process.

\section{Ionization stabilization}
\label{sec:IP}

\begin{figure}
 \includegraphics[width=0.5\textwidth]{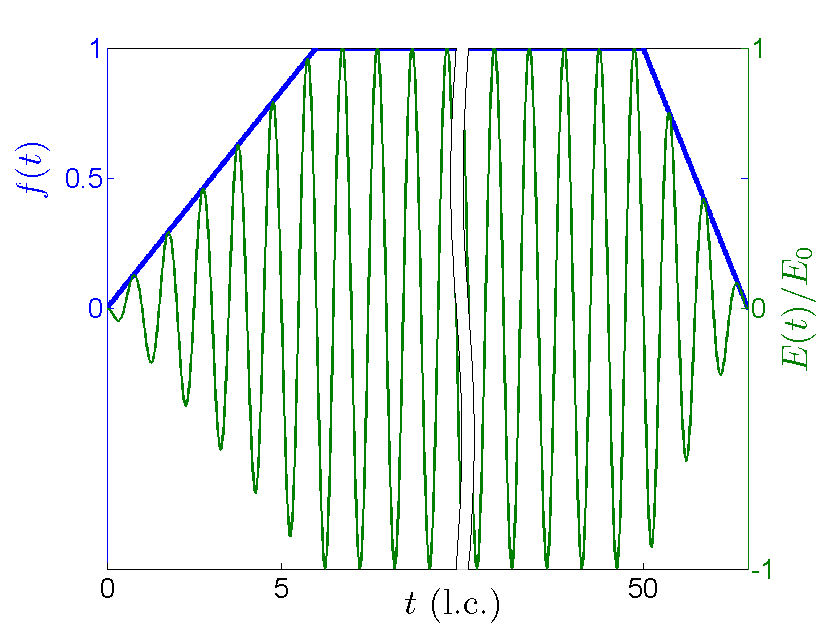}%
 \caption{\label{fig:pulse} Schematic representation of the pulse envelope $f(t)$ (left axis, bold line) and electric field $E(t)=-E_0f(t) \sin \omega t$ (right axis, thin line) used in Hamiltonian~(\ref{eq:Ham}). Since the plateau is much longer that the ramp-up and the ramp-down, it has not been fully depicted. }%
 \end{figure}

We consider the following classical model for a one-dimensional single active electron atom interacting with an intense linearly polarized laser field in the dipole approximation~:
\begin{equation}
\label{eq:Ham}
H(x,p,t)=\frac{p^2}{2}-\frac{1}{\sqrt{x^2+1}}-E_0 f(t) x\sin\omega t,
\end{equation}
where $p$ is the momentum of the electron canonically conjugate to the position $x$ along the polarization axis, $E_0$ is the amplitude of the electric field, $f(t)$ is the pulse envelope (schematically represented on Fig.~\ref{fig:pulse}) and $\omega$ is the frequency. We have chosen a soft Coulomb potential~\cite{java88,beck12} as the interaction potential between the electron and its parent ion. In Ref.~\cite{dziu10}, the impact of the choice of potential on stabilization has been analyzed. In all the computations presented in this article, the laser frequency is $\omega=0.8~{\rm a.u.}$ which corresponds to a wavelength of $57~{\rm nm}$.

\begin{figure}
\includegraphics[width=0.5\textwidth]{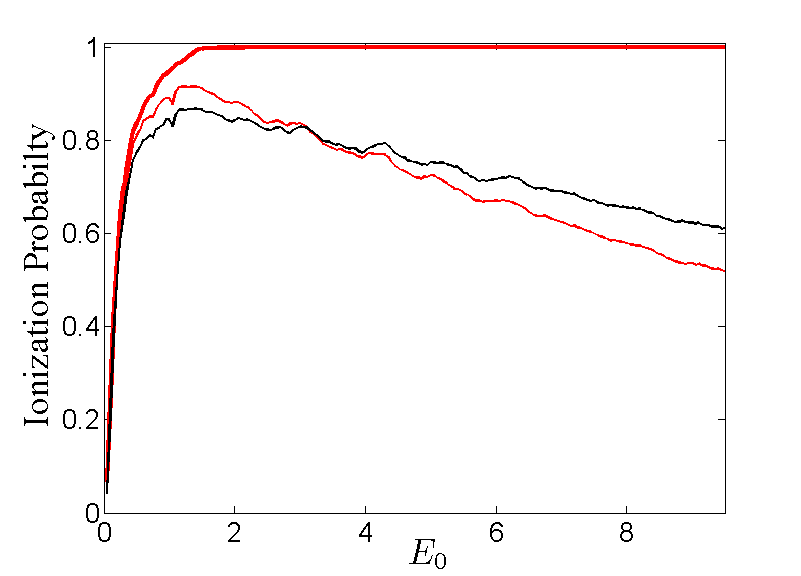}
 \caption{\label{fig:ioniz} Ionization probabilities as functions of $E_0$ obtained for Hamiltonian~(\ref{eq:Ham}) with a ramp-up of duration $T_{\rm u}=6T$, a plateau of duration $T_{\rm p}=44T$. The bold curve (red online) is without a ramp-down ($T_{\rm d}=0$) and using the energy criterion for ionization. The thin gray curve (red online) is without a ramp-down ($T_{\rm d}=0$), using a distance criterion with a threshold of $50~{\rm a.u.}$. The black curve is with a ramp-down of duration $T_{\rm d}=6T$ and using the energy criterion.  } 
\end{figure}

We begin with revisiting a numerical experiment performed in Ref.~\cite{grob91}. For the envelope (see Fig.~\ref{fig:pulse}), we select a smooth turn-on consisting of a linear ramp-up of duration $T_{\rm u}=6T$, a plateau of $T_{\rm p}=44 T$, where $T=2\pi/\omega$ is the period of the laser, and an abrupt ramp-down, i.e., of duration $T_{\rm d}=0$. We consider a large ensemble of trajectories which are initially bounded, i.e., whose energy, defined as the sum of the kinetic energy plus the soft Coulomb potential, is negative. Typically for a given $E_0$ the number of initial conditions we consider is on the order of several millions in order to have good statistics. As a function of the amplitude $E_0$, we measure the ionization probability where ionization is determined using the energy criterion as in Ref.~\cite{grob91}.  
In Fig.~\ref{fig:ioniz}, the ionization probability is represented as a function of the amplitude of the electric field $E_0$ (bold gray curve, red online). Contrary to Ref.~\cite{grob91} we observe a roughly monotonic increase of the ionization probability with $E_0$ as it could be naively guessed (the more intense the laser field is, the more ionization there is). Now, instead of using the energy criterion as ionization criterion, we use a distance criterion. This distance criterion appears to be more subjective since it depends on an arbitrarily chosen threshold. Here the chosen threshold is 50 a.u.; we have checked that the results are not changed qualitatively by changing this threshold significantly. The resulting ionization curve is depicted in Fig.~\ref{fig:ioniz} (thin gray curve, red online). Using this distance criterion we are able to reproduce, at least qualitatively, the surprising results obtained in Ref.~\cite{grob91}, namely the lack of complete ionization at very high intensity and the global decrease of the ionization probability as the intensity of the laser field is increased (see Fig.~\ref{fig:ioniz}).
Here we first notice the importance of the ionization criterion to observe the ionization stabilization phenomenon. We performed a third numerical experiment by including a ramp-down with $T_{\rm d}=6T$ and by using the energy criterion. Contrary to the case with an abrupt ramp-down ($T_{\rm d}=0$), the energy criterion leads to stabilization (see Fig.~\ref{fig:ioniz}, black curve).   
From these numerical simulations, it appears that the energy criterion can only be used in the presence of a ramp-down, whereas the distance criterion always displays the stabilization phenomenon. In what follows, we only use the distance criterion unless specified otherwise. Below we provide an explanation for the differences resulting from the two choices of ionization criterion. This is a direct result of the mechanism by which an electron remains trapped near the parent ion at a very high intensity, a mechanism we unravel using nonlinear dynamics.

\section{Analysis of stabilization from the nonlinear dynamics perspective}
\label{sec:NLanalysis}

\subsection{Role of the plateau}

We first elucidate the role of the plateau by providing the scenario for the lack of complete ionization at high intensity. Since the amplitude of the electric field is large, we consider first the Strong Field Approximation (SFA) which consists of neglecting the soft Coulomb interaction between the electron and the parent ion. In SFA, Hamiltonian~(\ref{eq:Ham}) during the plateau becomes
\begin{equation}
\label{eq:Ham0}
H_{\rm SFA}(x,p,t)=\frac{p^2}{2}-E_0 x \sin (\omega t+\phi),
\end{equation}
where we have included a laser phase to take into account the ramp-up, i.e., we consider that the plateau starts at $t=0$. This Hamiltonian is integrable and the trajectories can be explicitly provided~:
\begin{eqnarray*}
&& p(t)=p_0-\frac{E_0}{\omega}\left( \cos(\omega t+\phi)-\cos\phi\right),\\
&& x(t)=x_0+\left(p_0+\frac{E_0}{\omega}\cos\phi\right)t-\frac{E_0}{\omega^2}\left( \sin(\omega t+\phi)-\sin\phi\right),
\end{eqnarray*}
where $x_0$ and $p_0$ are the initial position and initial momentum (at $t=0$, i.e., at the beginning of the plateau). 
We notice a linear drift with velocity $v_d=p_0+(E_0/\omega)\cos\phi$. If this drift velocity is non-zero, the electron will eventually move far away from the parent ion, provided that the plateau is sufficiently long. We notice a very special set of initial conditions for which the drift velocity vanishes. This set in phase space is composed of an infinite family of (parabolic) periodic orbits with the same period as the laser field, related to each other by translation along the $x$-axis due to the continuous symmetry of Hamiltonian~(\ref{eq:Ham0}). More specifically, these periodic orbits have the equations
\begin{subequations}
\label{eq:POsfa}
\begin{eqnarray}
&& p(t)=-\frac{E_0}{\omega}\cos(\omega t+\phi),\\
&& x(t)=x_1-\frac{E_0}{\omega^2}\sin(\omega t+\phi),
\end{eqnarray}
\end{subequations}
where $x_1$ is an arbitrary position, the position of the center of the periodic orbit. If the electron is on one of these orbits at the beginning of the plateau, it will stay on it for an infinite amount of time. These invariant structures are the keystone to understand the lack of complete ionization at high intensity. 

Even though this set of periodic orbits is of measure zero and therefore it could be argued that it does not contribute significantly to the ionization probability, the nonlinearity introduced by the soft Coulomb interaction provides the reasons for the importance of this set of periodic orbits in the stabilization process, as we show next.  
  
The question is what happens to the rather simple SFA picture of the dynamics when the Coulomb field is taken into account? As the effect of the soft Coulomb potential increases, e.g., by increasing an effective charge up to 1, most of the periodic orbits~(\ref{eq:POsfa}) are broken, except a finite number of them. Three periodic orbits are of particular interest and organize the dynamics. All three periodic orbits have the same period, the period of the laser field. One of them is symmetric with respect to the $x=0$ axis, centered at $(x,p)=(0,0)$ and weakly hyperbolic. This periodic orbit is denoted by ${\cal O}$. The other two periodic orbits of interest are elliptic and symmetric with each other, centered approximately at $(x,p)=(\pm E_0/\omega^2,0)$. These orbits are denoted ${\cal O}_\pm$. The property of a periodic orbit to be hyperbolic, parabolic or elliptic features refers to the linear stability analysis (see Ref.~\cite{Bchaos} for more detail). In general, an elliptic periodic orbit is stable and the motion around it is similar to the motion around a stable equilibrium point of a simple pendulum, whereas an hyperbolic orbit is (linearly) unstable, in the sense that almost all small perturbation around it will drive the motion away from it (along its unstable manifold). The chaotic behavior of the system originates in the neighborhood of hyperbolic periodic orbits. 

\begin{figure}
\includegraphics[width=0.5\textwidth]{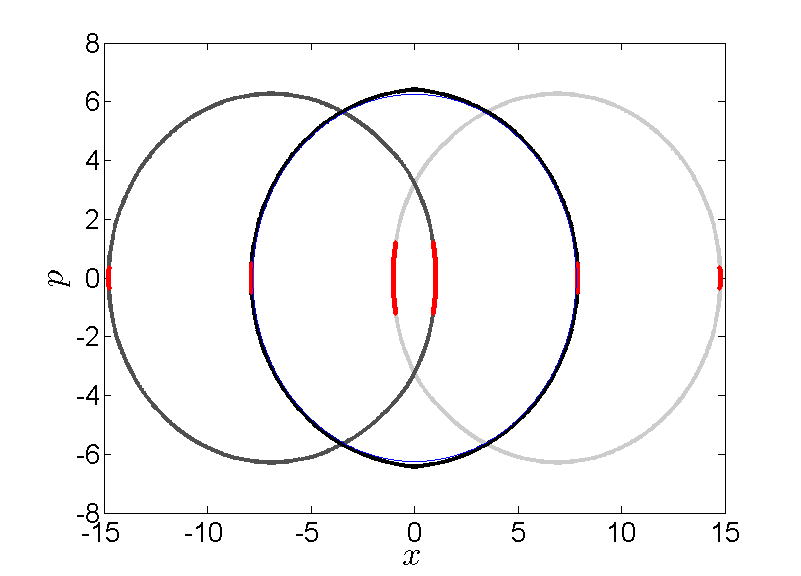}
 \caption{\label{fig:PO_O} Projection of the three periodic orbits ${\mathcal O}$ and ${\cal O}_\pm$ of Hamiltonian~(\ref{eq:Ham}) with $f(t)=1$ and $E_0=5$. The light gray orbit is ${\cal O}_+$, the darker gray orbit is ${\cal O}_-$ and ${\cal O}$ is represented in black. The gray parts (red online) of the orbits are where the energy (defined as the sum of the kinetic energy and soft Coulomb potential) is negative. The thin line corresponds to the periodic orbit in the SFA approximation given by Eq.~(\ref{eq:POsfa}), almost indistinguishable from ${\cal O}$. } 
\end{figure}

\begin{figure}
\includegraphics[width=0.5\textwidth]{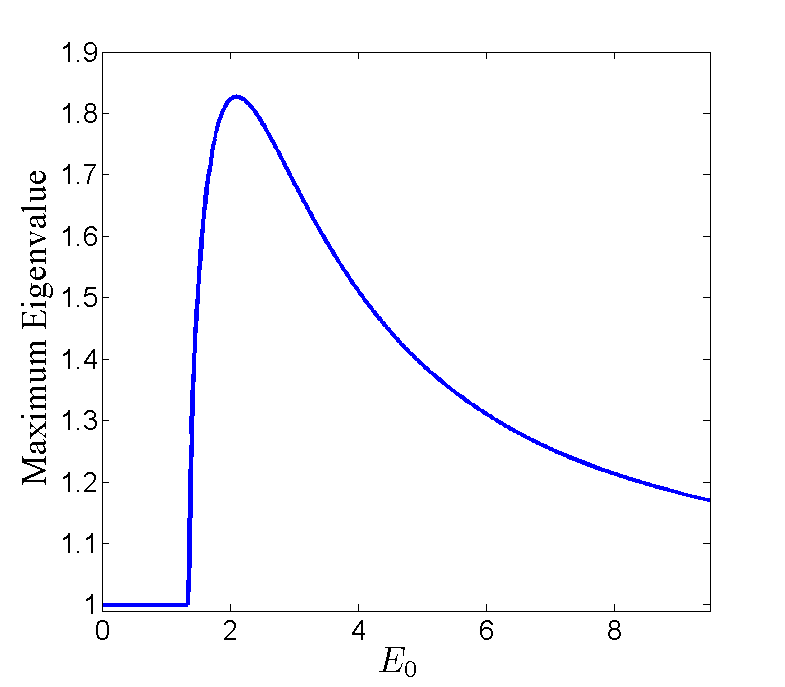}
 \caption{\label{fig:eigen} Largest eigenvalue of the periodic orbit ${\mathcal O}$ (represented in Fig.~\ref{fig:PO_O}) of Hamiltonian~(\ref{eq:Ham}) as a function of $E_0$.  } 
\end{figure}

\begin{figure}
\includegraphics[width=0.5\textwidth]{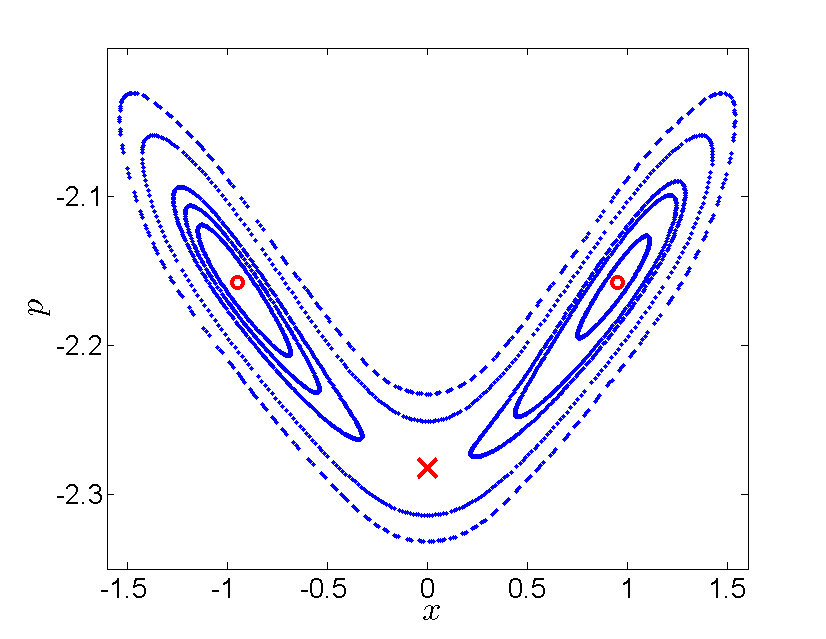}
 \caption{\label{fig:PS15} Poincar\'e section (stroboscopic plot with period $T$) of trajectories of Hamiltonian~(\ref{eq:Ham}) for $E_0=1.5$. The positions of the periodic orbits are indicated by a cross for ${\cal O}$ and by circles for the elliptic periodic orbits ${\cal O}_\pm$. } 
\end{figure}

\begin{figure}
\includegraphics[width=0.5\textwidth]{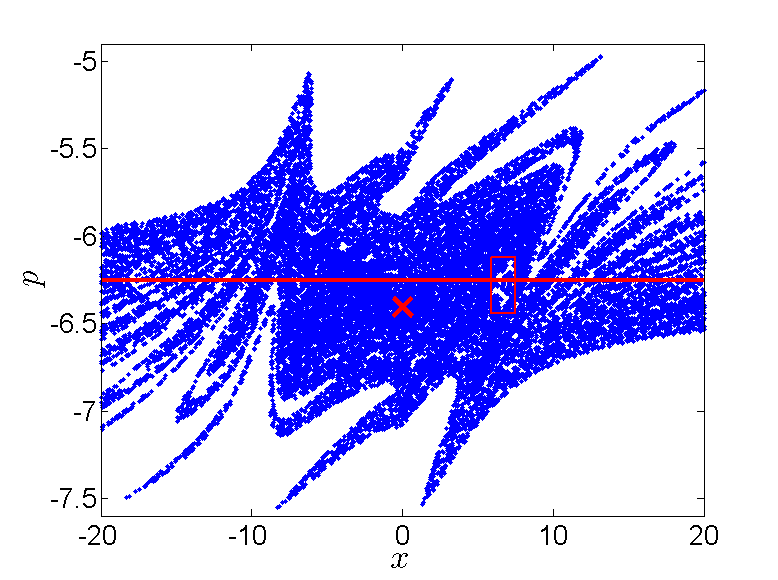}
 \caption{\label{fig:PS} Poincar\'e section of trajectories of Hamiltonian~(\ref{eq:Ham}) for $E_0=5$ in the region close to $p_0=-E_0/\omega$ (indicated by a continuous horizontal line). The position of the periodic orbit $\cal O$ is indicated by a cross. The box represents the inset where the Poincar\'e section is depicted in Fig.~\ref{fig:PSinset}.  } 
\end{figure}

\begin{figure}
\includegraphics[width=0.5\textwidth]{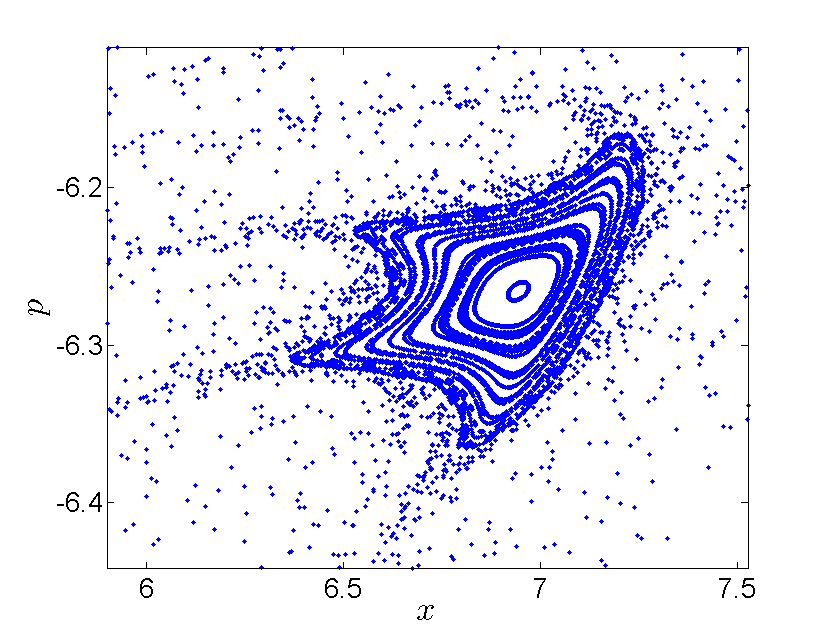}
 \caption{\label{fig:PSinset} Inset of Fig.~\ref{fig:PS}.  } 
\end{figure}

\begin{figure}
\includegraphics[width=0.5\textwidth]{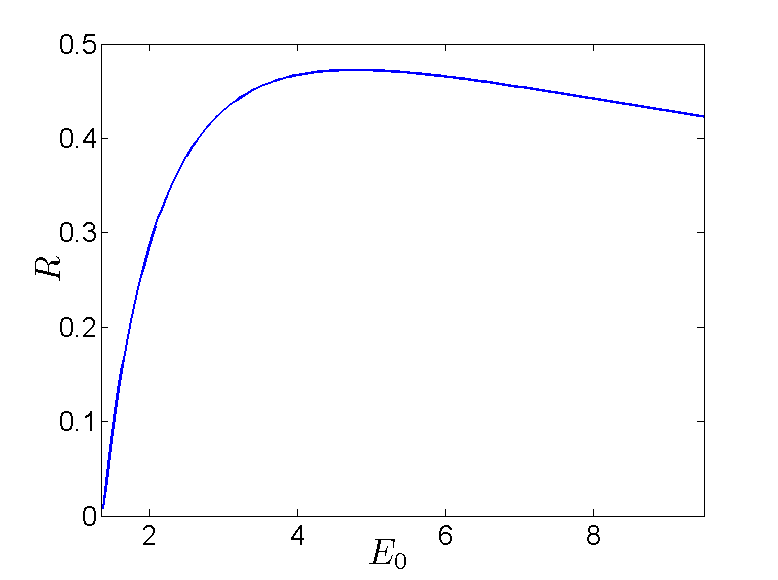}
 \caption{\label{fig:residue} Greene's residue $R$ of ${\cal O}_+$  (or equivalently ${\cal O}_-$) as a function of $E_0$.  } 
\end{figure}

Figure~\ref{fig:PO_O} displays the three periodic orbits of Hamiltonian~(\ref{eq:Ham}) with $f(t)=1$ and for $E_0=5$. At the high values of the intensity we consider in this paper, ${\cal O}$ is very close to the periodic orbit given by Eqs.~(\ref{eq:POsfa}) for $x_1=0$ (see Fig.~\ref{fig:PO_O} where it is difficult to distinguish the orbit~(\ref{eq:POsfa}) from the orbit ${\cal O}$). In Fig.~\ref{fig:eigen} we represent the maximum eigenvalue of the monodromy matrix associated with ${\cal O}$ as a function of $E_0$ (see Ref.~\cite{Bchaos} for more details). We notice that for $E_0$ small, ${\cal O}$ is elliptic and, as $E_0$ is increased, undergoes a bifurcation around $E_0=1.34$. The orbit ${\mathcal O}$ turns hyperbolic and two asymmetric and elliptic periodic orbits are created, these being ${\cal O}_\pm$. The basic structure of phase space is a horizontal 8-shape with two symmetric elliptic islands with the orbits ${\cal O}_\pm$ in their centers and the orbit ${\cal O}$ at the crossing. A Poincar\'e section of selected trajectories consisting of a stroboscopic plot of period $T$ is represented on Fig.~\ref{fig:PS15} for $E_0=1.5$, i.e., close to the bifurcation. This picture persists at higher value of $E_0$, although complicated by the creation of a dense chaotic tangle around these orbits. A Poincar\'e section of the region close to the weakly periodic orbit is depicted in Fig.~\ref{fig:PS} for $E_0=5$. The dense chaotic region is clearly observed, extending to high values of the position of the electron. Trajectories starting in the white region ionize very quickly, contrary to the trajectories started in the blue region which can be potentially trapped for an arbitrary long time. The two symmetric elliptic islands are present at this very high value of the intensity (see Fig.~\ref{fig:PSinset}), demonstrating the stabilization effect caused by the absence of drift velocity in the SFA approximation. In this chaotic region, there are many other periodic orbits and some of them might even be elliptic (depending on $E_0$) forming islands of stability (trapping trajectories at all times).
It should be noted that the two periodic orbits ${\cal O}_\pm$ remains elliptic for all values of $E_0$, becoming closer and closer to parabolic as $E_0$ is increased. In Fig.~\ref{fig:residue}, we represent Greene's residue~\cite{gree79,mack92} of ${\cal O}_\pm$ as a function of $E_0$. We recall that that if the residue is between 0 and 1, the periodic orbit is elliptic. This property is very interesting since the excited states around ${\cal O}_\pm$ (which extends up to the quiver radius, see Fig.~\ref{fig:PO_O}) are stable states.  

In a nutshell, the effect of the soft Coulomb interaction is to create a chaotic layer which strongly affects the transport properties of the system, trapping trajectories for a sufficiently long time compared to the duration of the plateau. These trajectories will not ionize, in the sense that they will not significantly depart from a region close to the core, contrary to the trajectories experiencing a drift velocity.
This chaotic region explains the lack of complete ionization, even at very high values of the laser intensity. 
Now, concerning the results obtained using the energy criterion, we notice that by looking at the energy of the non-ionizing trajectories (where non-ionizing feature has been determined using a distance criterion), given the values of the momentum which extend up to $E_0/\omega$, most of these trajectories have positive energy. For instance, we have displayed in red on Fig.~\ref{fig:PO_O} the points of the orbits ${\cal O}$ and ${\cal O}_\pm$ which have negative energy (and would be considered as non-ionizing using an energy criterion).
Therefore, according to an energy criterion, most of the trapped trajectories are considered as ionizing even though they are of a very different nature than the ones experiencing a drift velocity.  
 
\begin{figure}
\includegraphics[width=0.5\textwidth]{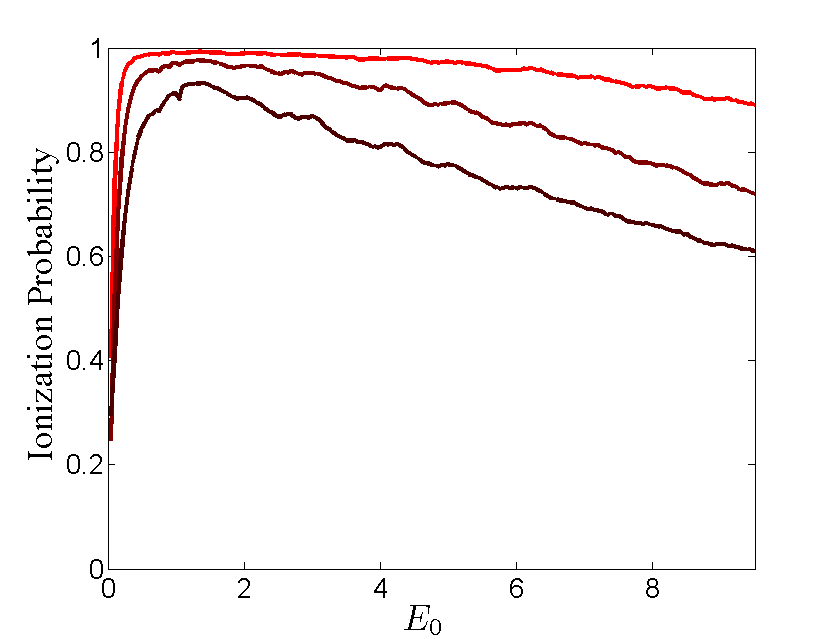}
 \caption{\label{fig:IP_plateau} Ionization probabilities as functions of $E_0$ for different durations of the plateau. From black to lighter gray (red online), the total duration of the pulse is 50, 100 and 200 laser cycles, with a 6 laser cycle ramp-up and no ramp-down.  } 
\end{figure}

Globally the effect of the plateau is to discriminate between the trajectories with a drift velocity with the ones trapped in the chaotic tangle. The longer the pulse is, the more trapped trajectories leave the chaotic tangle, and eventually these will be considered as ionized (even using a distance criterion). In Fig.~\ref{fig:IP_plateau}, we represent the ionization probabilities as functions of $E_0$ for different values of the plateau duration. It confirms that the ionization probability increases with the plateau duration. However it should be noticed that due to the presence of elliptic islands, the ionization will never be complete for most values of the laser intensity, since the trajectories in these regions are trapped for all times, preventing the ionization probability to reach 1 regardless of the length of the plateau.  
It should be noticed that the finer structure of the ionization curve, like the oscillations, are qualitatively similar (located at approximately the same values of $E_0$), regardless of the duration of the plateau. 

\begin{figure}
\includegraphics[width=0.5\textwidth]{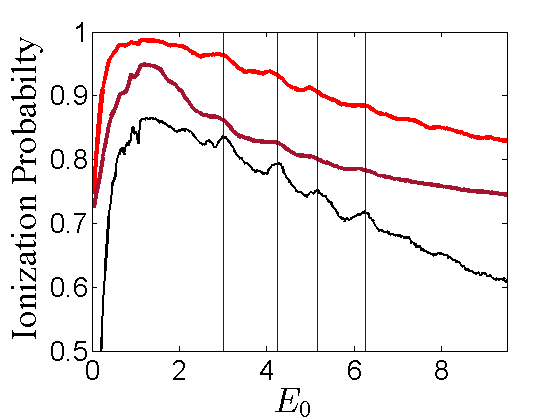}
 \caption{\label{fig:sizeChaos} Probability of a trajectory of Hamiltonian~(\ref{eq:Ham}) with no ramp-up and no ramp-down to leave the window $(x,p)\in [-20,20]\times[-E_0/\omega-3/2,-E_0/\omega+3/2]$ after 20 (light gray, red online) and 50 (dark gray) laser cycles. For comparison, we depict the ionization curve of Fig.~\ref{fig:ioniz} obtained with a ramp-up of duration $T_{\rm u}=6T$ and no ramp-down (black curve). } 
\end{figure}

Next we look at the size of the chaotic/trapping region as a function of $E_0$. In order to do this, we launch a high number of trajectories in a window around the periodic orbit in the Poincar\'e section. The window we consider is $(x,p)\in [-20,20]\times[-E_0/\omega-3/2,-E_0/\omega+3/2]$, where we recall that the periodic orbit ${\cal O}$ is locate approximately at $(0,-E_0/\omega)$. We look at trajectories which have left this window after 20 and 50 laser cycles. In Fig.~\ref{fig:sizeChaos}, we represent the probability of a trajectory to leave the window of interest. We compare this probability with the ionization probability of Fig.~\ref{fig:ioniz} with a ramp-up of duration $T_{\rm u}=6T$ and no ramp-down. The probability to leave the window exhibits the main features of the ionization probability. First it is decreasing (meaning that the size of the chaotic region increases with $E_0$), therefore displaying the stabilization process. In addition, we notice that this curve displays the same finer structures (the oscillations as function of $E_0$). Therefore the structure of the ionization curves of Fig.~\ref{fig:ioniz} is a nonlinear effect caused by the presence of both the soft Coulomb potential and the laser field. As expected the ionization probability is smaller due to the fact that the average intensity felt by the electron is smaller when there is a ramp-up. The ramp-up has also another impact on the ionization probability and this will be analyzed below.   
The global decrease of the ionization curves is due to the fact that, as $E_0$ is increased the system becomes closer to integrable as shown in Figs.~\ref{fig:eigen} and \ref{fig:residue} where the periodic orbits becomes closer to parabolic.  

It should be noted that the same periodic orbits ${\cal O}$ and ${\cal O}_\pm$ which control the stabilization process, also control the recollision process by allowing an electron far away from the parent ion to come back to it and recollide. These periodic orbits were coined recolliding periodic orbits (RPO) in Refs.~\cite{kamo13,kamo14}.  
 
\subsection{Role of the ramp-up}

\begin{figure}
\includegraphics[width=0.5\textwidth]{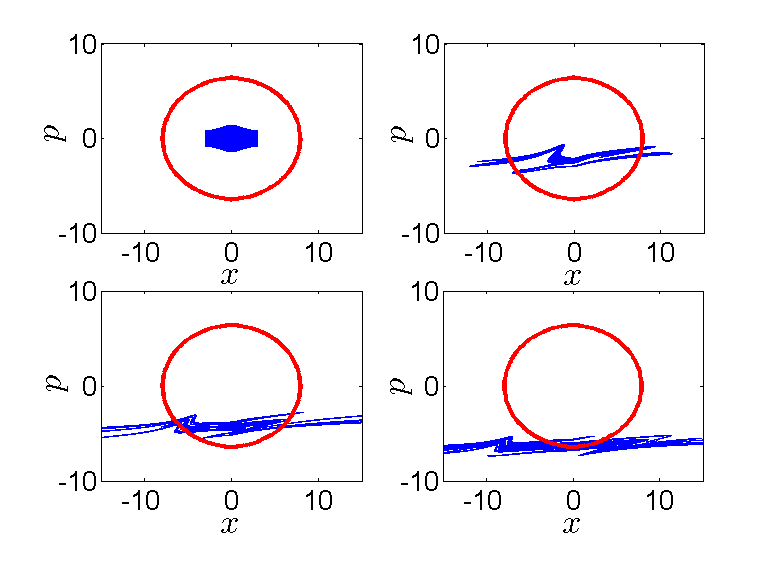}
 \caption{\label{fig:promote} Positions of an ensemble of trajectories of Hamiltonian~(\ref{eq:Ham}) for $E_0=5$ at different stages of the ramp-up, $t=0$ (upper left panel), $t=2T$ (upper right panel), $t=4T$ (lower left panel), and at the end of the ramp-up $t=6T$ (lower right panel). The periodic orbit ${\cal O}$ is depicted in gray (red online).} 
\end{figure}

The role of the ramp-up is important in setting up the right conditions for some of the trajectories to be put inside the chaotic/trapping region at the beginning of the plateau. In order to illustrate this argument, we first consider the SFA approximation, by considering the following Hamiltonian~:
$$
H_{\rm u}(x,p,t)=\frac{p^2}{2}-E_0 x\frac{t}{T_{\rm u}}\sin (\omega t+\phi),
$$
where here the phase $\phi$ is used to take into account the transient time when the SFA approximation is not valid. The momentum and the position at the end of the ramp-up are
\begin{eqnarray*}
&& p_{\rm u}=p_0-\frac{E_0}{\omega}\left(\cos(\omega T_{\rm u}+\phi)-\frac{\sin(\omega T_{\rm u}+\phi)-\sin\phi}{\omega T_{\rm u}}\right),\\
&& x_{\rm u}=x_0-\frac{E_0}{\omega^2}\left( \sin(\omega T_{\rm u}+\phi)+\sin\phi 
\right. \\
&&\qquad \qquad \qquad \left. 
+2\frac{\cos(\omega T_{\rm u}+\phi)-\cos\phi}{\omega T_{\rm u}}\right).
\end{eqnarray*}
If the duration of the ramp-down is an integer multiple of the laser period then it reduces to
\begin{eqnarray*}
&& p_{\rm u}=p_0-\frac{E_0}{\omega}\cos\phi,\\
&& x_{\rm u}=x_0-2\frac{E_0}{\omega^2}\sin\phi,
\end{eqnarray*}
which is a point on a periodic orbit given by Eqs.~(\ref{eq:POsfa}) with $x_1=x_0-(E_0/\omega^2)\sin\phi$, provided that $p_0=0$. If the initial distribution at the beginning of the ramp-up is close to zero then the role of the ramp-up (at least in the SFA approximation) is to promote these initial conditions to a region where periodic orbits organize the dynamics and prevent ionization. 
\begin{figure}
\includegraphics[width=0.5\textwidth]{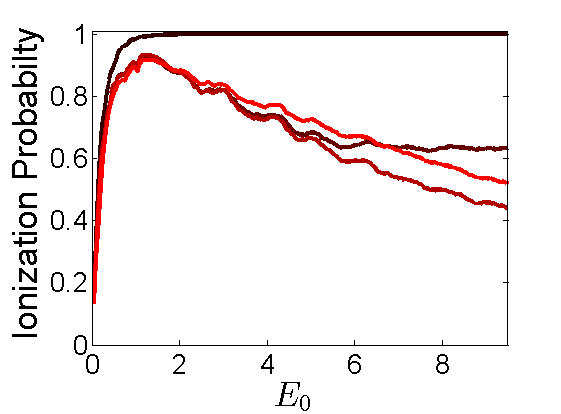}
 \caption{\label{fig:IP_up} Ionization probabilities as functions of $E_0$ for different durations of the ramp-up. From black to lighter gray (red online), the duration of the ramp-up is $T/10$, $T/2$, $T$ and $6T$ with a 44 laser cycle plateau and no ramp-down.  } 
\end{figure}
Taking into account the soft Coulomb potential, the role of the ramp-up is to promote some of the trajectories to the chaotic/trapping region organized by the three periodic orbits ${\cal O}$ and ${\cal O}_\pm$. In order to demonstrate this effect of the ramp-up, we depict the position of an ensemble of electrons in phase space at different stages in the ramp-up. Initially these trajectories are bounded, in the sense that their energy (sum of kinetic energy and soft Coulomb potential is negative). In Fig.~\ref{fig:promote}, four stages are represented~: at $t=0$, $t=2T$, $t=4T$ and $t=6T$. We notice that initially none of the trajectories are on the periodic orbit. However, as early as half of the ramp-up, some of the trajectories are already on the periodic orbit or nearby. 
If the ramp-up is not sufficient to promote the trajectories to the chaotic/trapping region, for instance, if it is too short (shorter than half a laser cycle for instance), then all the trajectories will experience a significant drift velocity and will be all ionized. Therefore the ramp-up does not play a role in the stabilization mechanism but allows part of the initial conditions to undergo this stabilization. 

\subsection{Role of the ramp-down}

Using the energy criterion, the presence of the ramp-down seems essential for the stabilization. However we have shown above that the mechanism for stabilization has nothing to do with the ramp-down. In this section we analyze the effect of the ramp-down, showing that it only plays a minor role in the dynamics. However it is essential if one wants to use the energy criterion for ionization. 
In order to illustrate this, we consider the effect of the ramp-down on the prototypical trapped trajectory in the SFA approximation, namely the periodic orbit given by Eqs.~(\ref{eq:POsfa}). We shift the origin of time, so that the ramp-down starts at $t=0$ (and we assume that the ramp-up and plateau are integer multiples of the laser period). The duration of the ramp-down is $T_{\rm d}$. We consider the following Hamiltonian in the SFA approximation during the linear ramp-down:
$$
H_{\rm d}(x,p,t)=\frac{p^2}{2}-E_0 x \frac{T_{\rm d}-t}{T_{\rm d}}\sin(\omega t+\phi).
$$
At $t=0$ (i.e., at the end of the plateau), we assume that the trajectory is on the periodic orbit, i.e., the initial condition is $x_0=-(E_0/\omega^2)\sin\phi$ and $p_0=-(E_0/\omega)\cos\phi$ at the beginning of the ramp-down. At the end of the ramp-down, the final momentum is
$$ 
p_{f}=\frac{E_0}{\omega}\frac{\sin\phi-\sin(\omega T_{\rm d}+\phi)}{\omega T_{\rm d}}.
$$
If the duration of the ramp-up is an integer multiple of $T$, then the final momentum vanishes. Regardless of the distance of the electron to the parent ion, its energy is then negative. Therefore the electron located on the SFA periodic orbit at the end of the plateau is considered ionized at the end of the ramp-down according to the energy criterion. However if the duration of the ramp-down is not an integer multiple of $T$, the final momentum can be large (up to $E_0/\omega$) and consequently can be considered as ionizing according to the energy criterion.  
It should be noted that using the distance criterion, the effect of the ramp-down can be neglected; all the ionization probability curves are nearly identical.  

\begin{figure}
\includegraphics[width=0.5\textwidth]{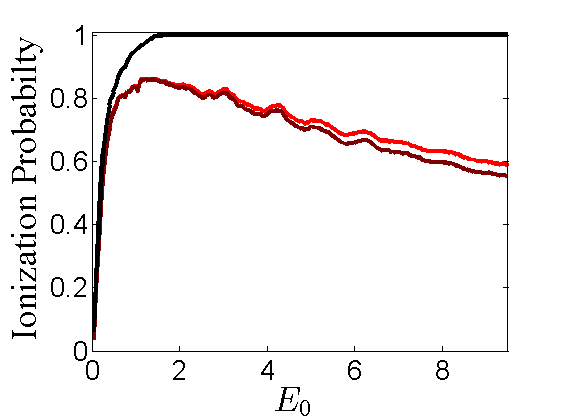}
 \caption{\label{fig:IP_down} Ionization probabilities as functions of $E_0$ obtained from Hamiltonian~(\ref{eq:Ham}) for different durations of the ramp-down. From black to lighter gray (red online), the duration of the ramp-down is $T_{\rm d}=T/5$ (black curve), $T_{\rm d}=T$ (dark gray, maroon online) and $T_{\rm d}=3T$ (light gray, red online), with a 6 laser cycle ramp-up and a 44 laser cycle plateau. The energy criterion is used to determine ionization.   } 
\end{figure}

In Fig.~\ref{fig:IP_down} we represent the ionization probabilities obtained with an energy criterion for different durations of the ramp-down, one of them being not an integer multiple of the laser period. Provided that the duration of the ramp-up is an integer multiple of the laser period, the two curves at $T_{\rm d}=T$ and $T_{\rm d}=3T$ demonstrate little effect of the ramp-down, even using the energy criterion as ionizing criterion.  

\section{Conclusion}
We have shown that the main mechanism for stabilization of atoms in a strong laser field is driven by a set of periodic orbits. The counter-intuitive nature of this phenomenon can be explained by the fact that the orbits mostly lay outside the bounded region.
We have elucidated the role of each phase in the laser pulse in the stabilization process. During the turn-on of the laser field, when the Coulomb and laser fields are of comparable strength, an electron is pushed towards a dense chaotic tangle in the neighborhood of these periodic orbits, trapping some trajectories for an arbitrarily long time (even for infinite time depending on the initial conditions). The role of the plateau is to seal the fate of the electron by discriminating the ones that are non-ionizing from the ionizing trajectories. We have showed that the role of the ramp-down is rather minor, allowing the use of the energy criterion, whereas, in the absence a ramp-down a distance criterion has to be used to detect the stabilization phenomenon.

\begin{acknowledgments}
T.U. acknowledges R. Grobe for some clarifications. M.J.N and T.U. acknowledge funding from the NSF. The research leading to these results has received funding from the People Program (Marie Curie Actions) of the European Union’s Seventh Framework Program No.~FP7/2007-2013/ under REA Grant No.~294974. P.W. acknowledges support from the Scientific Research Base Development Program of the Beijing Municipal Colleges for Higher Education Talent Programs (No.~067135300100).  
\end{acknowledgments}

\end{document}